\documentclass{article}
\usepackage{arxiv}
\usepackage[utf8]{inputenc} 
\usepackage[T1]{fontenc}    
\usepackage{hyperref}       
\usepackage{url}            
\usepackage{booktabs}       
\usepackage{amsfonts}       
\usepackage{nicefrac}       
\usepackage{microtype}      
\usepackage{lipsum}
\usepackage{graphicx}
\graphicspath{ {./images/} }
\title{Sentinels of the Stream: Unleashing Large Language Models for Dynamic Packet Classification in Software Defined Networks - Position Paper}
\author{
 Shariq Murtuza \\
  Department of Computer Science and Engineering \& Information Technology\\
  Jaypee Institute of Information Technology\\
  Noida, India \\
  \texttt{shariq.murtuza@jiit.ac.in} \\
}
\begin{document}
\maketitle
\begin{figure}[h]
\centering
\includegraphics[width=0.3\linewidth]{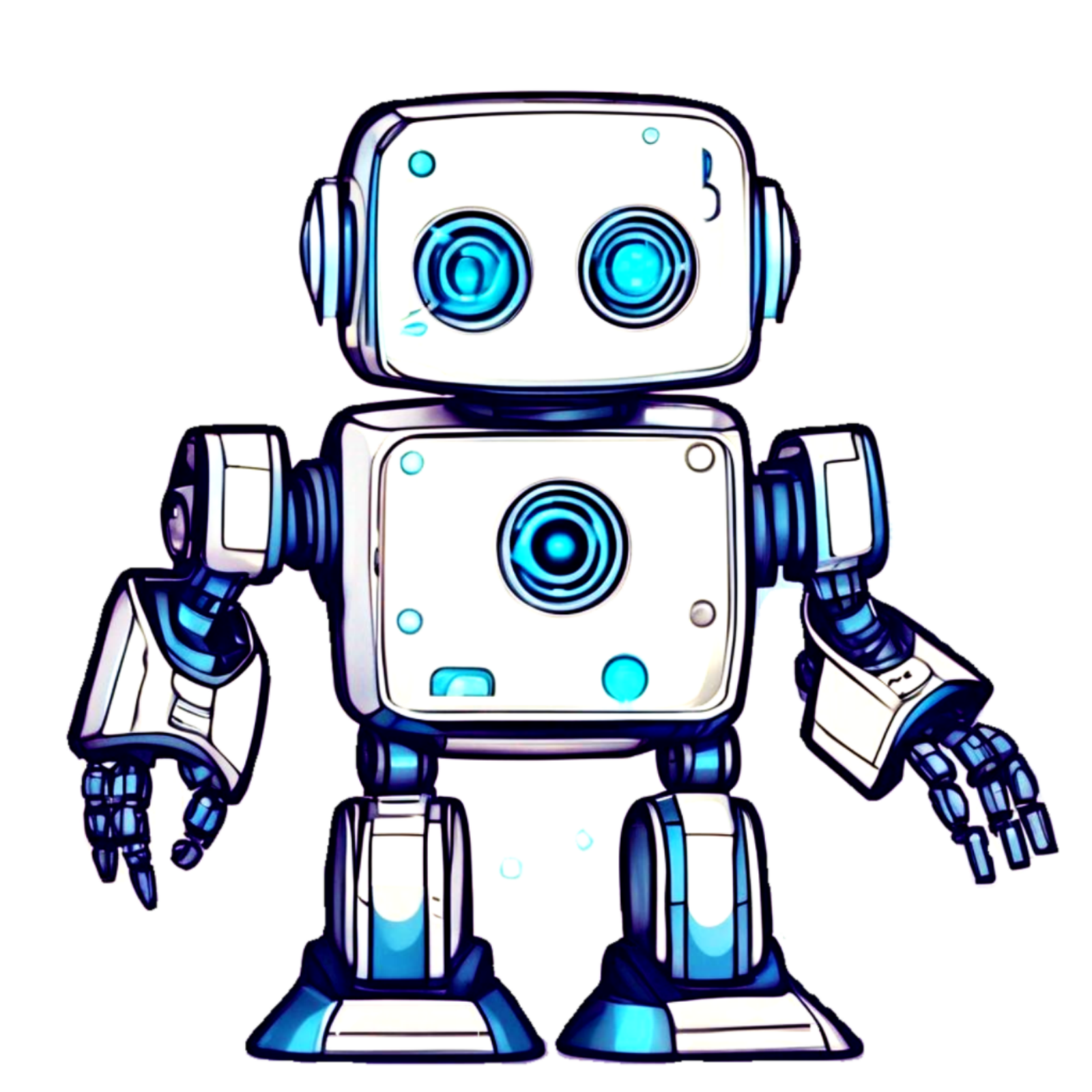}
\end{figure}
\begin{abstract}
With the release of OpenAI's ChatGPT, the field of large language models (LLM) saw an increase of academic interest in GPT based chat assistants. In the next few months multiple accesible large language models were released that included Meta's LLama models and Mistral AI's Mistral and Mixtral MoE models. These models are  available openly for a wide array of purposes with a wide spectrum of licenses. These LLMs have found their use  in a different number of fields like  code development, SQL generation etc. In this work we  propose our plan to explore the applicability of large language model in the domain of network security. We plan to create Sentinel, a LLM, to analyse network packet contents and pass a judgment on it's threat level. This work is a preliminary report that will lay our plan for our future endeavors. \footnote{Technical Report, work in progress.}
\end{abstract}
\keywords{Large Language Models \and Network Security \and Distributed Denial of Service Attacks \and Traffic Classification}

\section{Introduction}
With the launch of ChatGPT \cite{1,2} the field of Artificial Intelligence experienced a renewed interest from all walks of life. ChatGPT is a type of language model called a Large Language Model (LLM) \cite{4}. In simple terms,  a LLM is initially trained on data using semi supervised and self supervised learning, and then produce output by predicting the next token.  LLMs have strong potential to perform   a wide range of tasks \cite{5,6} specified in typical spoken language or technically, in natural language. These models can extensively understand and perform Natural Language Processing (NLP) tasks such as writing  letters, invitation, apologies, participate in a dialogue, give solution to mathematical question and lot more. Even though the main expertise of these models is in language(pattern) based tasks, these models can also be used to produced highly specialised texts similar to humans \cite{7} based on the given instruction. A network packet is an ordered collection of bits with each bit reserved for a specific flag representing a value. Each network packet is made up of a sequence of zeros and ones (bits) with each bit location earmarked for a specific purpose. The original IP protocol specification was give in RFC 791 in the year 1981 \cite{8}. While the TCP protocol  specification was given in RFC 9293 \cite{9}. Over time these specifications have been updated over time but have provided a concrete foundation for network stack. As each bit location has a specific meaning, we can map it as a sentence of a hypothetical language that consists of bit stream with individual components like flags, port number, IP address analogous to words.
Meta(earlier known as Facebook) launched  Llama and Llama2 (two different iterations a single  model)  and associated weights to the research community in 2023 with different licenses \cite{10,11}. The model was named as LLaMA (Large Language Model Meta AI) \cite{12,13} and was released in  February 2023, and the second model called LLaMA-2 \cite{14}, in July 2023. LLaMA was released in four different model sizes 7, 13, 33 and 65 billion parameters, while LLaMA-2 was released in three model sizes 7, 13, and 70 billion parameters. The next update in the public LLM domain came with the release of Mistral AI's Mistral 7B \cite{15,16} (in September, 2023) that outperformed all the  available open models up to 13B parameters, at that time, as per the existing language and code  benchmarks \cite{17,18} Soon after the Falcon  family of models (7B, 40B, and 180B) were released by the Technology Innovation Institute (TII) \cite{19,21} in November, 2023 outperforming the Llama-2 models on different benchmarks. These models are also open and are available for personal and commercial usage under different licences \cite{20}. 
In this work we present  our plan to finetune and create a large language model named Sentinel, that can  analyses network packets and identify malicious attack packets from a given network flow. We discuss the details in the next section.
\section{Creating our Large Language Model - Sentinel}
Finetuning is an approach often deployed in LLMs to increase their accuracy at the cost of shrinking their output domain \cite{3,22,23}. For example, an LLM fine-tuned on medical science data will give better results compared to an LLM that has not been fine-tuned yet \cite{24,25}. However, finetuning also reduces the number of different domains an LLM can respond to. In layman's terms, finetuning involves adjusting the weights of an already existing pre-trained model to enhance its performance on specific types of problems. Our work is based on the work done in  PADEC \cite{26} and extending it to encompass multiple different attack scenarios. 
We plan to  fine-tune three different  models namely, Llama2-7B, Falcon-7B and Mixtral MoE (mixture of experts) \cite{30,29} on our custom-generated SDN link flood attack dataset. Our dataset was generated with  the Containernet \cite{31} simulator. We created multiple  nodes, attackers, switches, and the controller and captured the network traffic. Traditional datasets were avoided due to their lack of alignment with the rapid advancements in network attacks. We are not creating a new LLM model from scratch, but instead we will be using existing open models and train them upon our dataset.
\subsection{Dataset Processing}
To create the dataset, we simulated a multihost network with various nodes providing different services. Our dataset comprises captured network packets obtained using the TCPdump tool \cite{32}. 
The traffic capture commenced after the attack was initiated and had stabilized over time. The attack packets comprises of different attack categories, including volumetric attacks, protocol attacks, and other vulnerability-based attacks \cite{33}. 

\subsection{Finetuning}
\begin{figure}
\includegraphics[width=\textwidth]{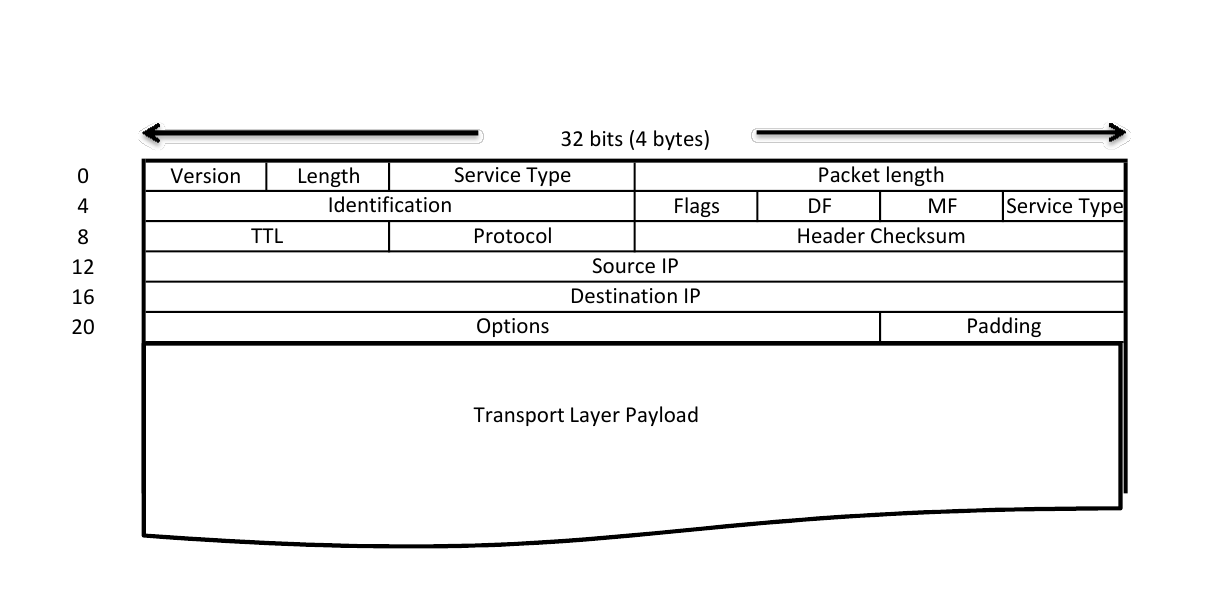}
\centering
\caption{IP Packet format}
\label{fig:ip}
\end{figure}

\begin{figure}
\includegraphics[width=\textwidth]{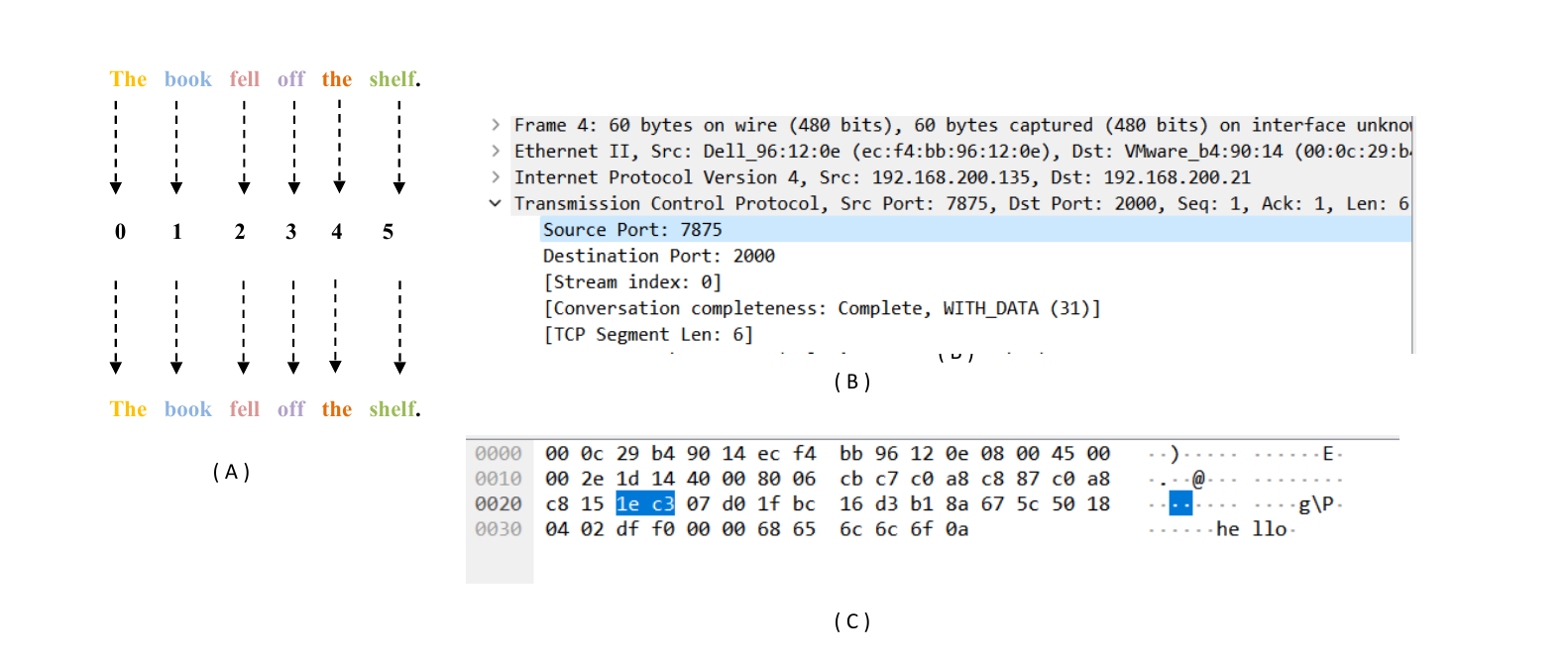}
\centering
\caption{Similarity of a sentence with an IP Packet}
\label{fig:wire}
\end{figure}

Each packet is a collection of bits, with each bit corresponding to a specific defined purpose in the network packet. This makes it very similar to natural spoken language, and thus we can safely assume a network packet to be a piece of textual language, where the language consists of only two alphabets: zero and one. Just as a language sentence has a format, a group of semantically related words delimited by special punctuation marks (comma, full stop, question mark, etc.), the network packets also have a clearly defined format, with each numbered bit reserved for a specific purpose. Similar to how the next word in a sentence are related to the current word, inside a network packet's header, the next bit(s) are related to the current bit value.
\begin{figure}
\includegraphics[width=\textwidth]{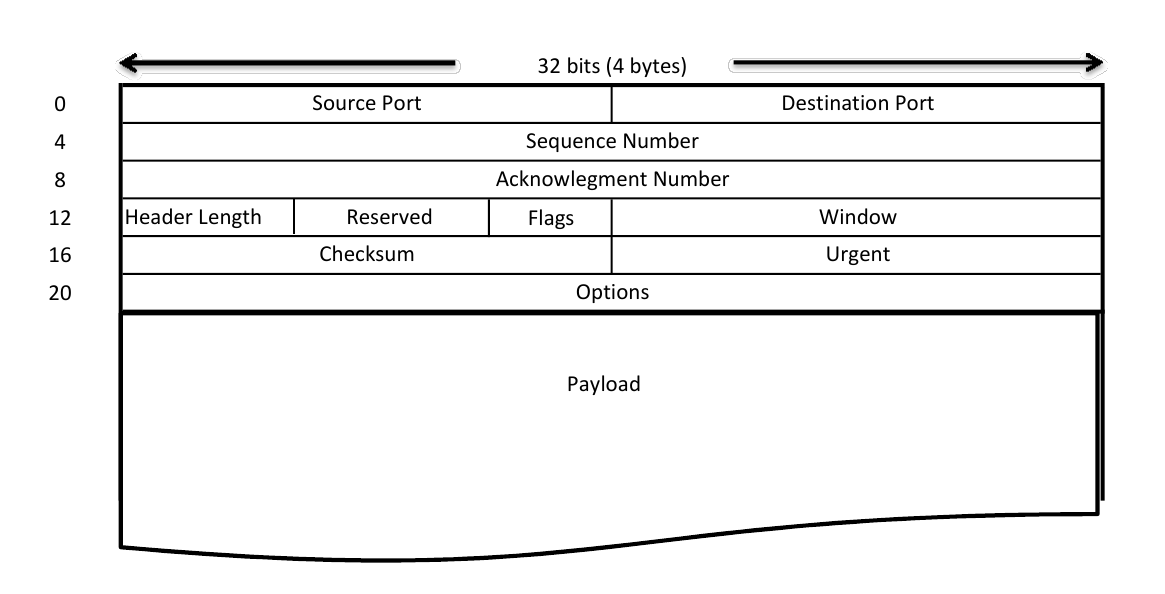}
\centering
\caption{TCP Packet format}
\label{fig:tcp}
\end{figure}
For example within the TCP header headers as shown in Fig. \ref{fig:tcp}, the source port is always followed by the destination port. Similarly an IP packet (shown in Fig. \ref{fig:ip} also follows specific rules and in no scenarios can we have a packet without an IP address and/or a port number. This compulsion makes a network packet highly similar to a language with grammar  rules and a captured network packet equivalent to a written sentence. As each bit location has a specific meaning, we can map it as a sentence of a hypothetical language that consists of bit stream with individual components like flags, port number, IP address analogous to words. As we have proven, a captured network packet is no different from a written sentence, thus we can safely say that a large language model can be created to read, write and  understand  a language that has it's constituents sentences in the form of network packets. The implied similarity is shown in Fig \ref{fig:wire} where it is evident that specific bits always refer to specific information in a packet. 
This   language (consisting of network packets) would be simpler than spoken languages like English due to the relative smaller possible vocabulary but in certain aspects it would be more complex since the different permutation and combinations of each flag in the header creates a new network packet variant.

\section{Conclusion}
In this text we share our plans to create Sentinel, a fine tuned LLM  to analyse network packets and take a decision regarding the amount of threat posed by the packets. We are starting with Llama2-7B then we will move towards Mistral AI's 7B model and shall finally study the feasibility of Mixtral MoE model. We are using three very different models to gauge their efficacy in handling this task and to select the best among the candidates. Our selection of model is purely on their performance on various public leader-boards and their proven capabilities in performing given tasks.
\bibliographystyle{unsrt}  

\begin{thebibliography}{1}
\bibitem{1} Achiam, Josh, Steven Adler, Sandhini Agarwal, Lama Ahmad, Ilge Akkaya, Florencia Leoni Aleman, Diogo Almeida et al. "Gpt-4 technical report." arXiv preprint arXiv:2303.08774 (2023).
\bibitem{2} Zhang, Min, and Juntao Li. "A commentary of GPT-3 in MIT Technology Review 2021." Fundamental Research 1, no. 6 (2021): 831-833.
\bibitem{3} Hu, Zhiqiang, Yihuai Lan, Lei Wang, Wanyu Xu, Ee-Peng Lim, Roy Ka-Wei Lee, Lidong Bing, and Soujanya Poria. "LLM-Adapters: An Adapter Family for Parameter-Efficient Fine-Tuning of Large Language Models." arXiv preprint arXiv:2304.01933 (2023).
\bibitem{4} Chang, Yupeng, Xu Wang, Jindong Wang, Yuan Wu, Linyi Yang, Kaijie Zhu, Hao Chen et al. "A survey on evaluation of large language models." ACM Transactions on Intelligent Systems and Technology (2023).
\bibitem{5} Chang, Yupeng, Xu Wang, Jindong Wang, Yuan Wu, Linyi Yang, Kaijie Zhu, Hao Chen et al. "A survey on evaluation of large language models." ACM Transactions on Intelligent Systems and Technology (2023).
\bibitem{6} Osco, Lucas Prado, Eduardo Lopes de Lemos, Wesley Nunes Gonçalves, Ana Paula Marques Ramos, and José Marcato Junior. "The Potential of Visual ChatGPT for Remote Sensing." Remote Sensing 15, no. 13 (2023): 3232.
\bibitem{7} Dou, Yao, Maxwell Forbes, Rik Koncel-Kedziorski, Noah A. Smith, and Yejin Choi. "Is GPT-3 text indistinguishable from human text? SCARECROW: A framework for scrutinizing machine text." arXiv preprint arXiv:2107.01294 (2021).
\bibitem{8} Postel, Jon. "Internet protocol―DARPA internet program protocol specification, RFC 791." (Internet protocol―DARPA internet program protocol specification) (1981).
\bibitem{9} Eddy, W., ed. "RFC 9293: Transmission Control Protocol (TCP)." (2022).
\bibitem{10} Llama. “Llama,” n.d. https://llama.meta.com/.
\bibitem{11} Touvron, Hugo, Thibaut Lavril, Gautier Izacard, Xavier Martinet, Marie-Anne Lachaux, Timothée Lacroix, Baptiste Rozière et al. "Llama: Open and efficient foundation language models." arXiv preprint arXiv:2302.13971 (2023).
\bibitem{12} Meta Research. “LLaMA: Open and Efficient Foundation Language Models - Meta Research | Meta Research,” February 24, 2023. https://research.facebook.com/publications/llama-open-and-efficient-foundation-language-models/.
\bibitem{13} Introducing LLaMA: A foundational, 65-billion-parameter language model. “Introducing LLaMA: A Foundational, 65-Billion-Parameter Language Model,” n.d. https://ai.meta.com/blog/large-language-model-llama-meta-ai/.
\bibitem{14} meta-llama (Meta Llama 2). “Meta-Llama (Meta Llama 2),” December 27, 2023. \url{https://huggingface.co/meta-llama}.
\bibitem{15} AI, Mistral. “Mistral AI | Open-Weight Models.” Mistral AI | Open-weight models, n.d. /.
\bibitem{16} AI, Mistral. “Mistral 7B.” Mistral 7B | Mistral AI | Open-weight models, September 27, 2023. \url{https://mistral.ai/news/announcing-mistral-7b/}.
\bibitem{17} Open LLM Leaderboard - a Hugging Face Space by HuggingFaceH4. “Open LLM Leaderboard - a Hugging Face Space by HuggingFaceH4,” n.d. \url{https://huggingface.co/spaces/HuggingFaceH4/open_llm_leaderboard.}
\bibitem{18} open-llm-leaderboard (Open LLM Leaderboard). “Open-Llm-Leaderboard (Open LLM Leaderboard),” February 6, 2024. \url{https://huggingface.co/open-llm-leaderboard}.
\bibitem{19} https://falconllm.tii.ae/. “Falcon LLM.” Accessed February 9, 2024. \url{https://falconllm.tii.ae/}.
\bibitem{20} tiiuae (Technology Innovation Institute). “Tiiuae (Technology Innovation Institute),” September 6, 2023. \url{https://huggingface.co/tiiuae}.
\bibitem{21} Penedo, Guilherme, Quentin Malartic, Daniel Hesslow, Ruxandra Cojocaru, Alessandro Cappelli, Hamza Alobeidli, Baptiste Pannier, Ebtesam Almazrouei, and Julien Launay. "The RefinedWeb dataset for Falcon LLM: outperforming curated corpora with web data, and web data only." arXiv preprint arXiv:2306.01116 (2023).
\bibitem{22} Peng, Baolin, Chunyuan Li, Pengcheng He, Michel Galley, and Jianfeng Gao. "Instruction tuning with gpt-4." arXiv preprint arXiv:2304.03277 (2023).
\bibitem{23} Ding, Ning, Yujia Qin, Guang Yang, Fuchao Wei, Zonghan Yang, Yusheng Su, Shengding Hu et al. "Parameter-efficient fine-tuning of large-scale pre-trained language models." Nature Machine Intelligence 5, no. 3 (2023): 220-235.
\bibitem{24} Thirunavukarasu, Arun James, Darren Shu Jeng Ting, Kabilan Elangovan, Laura Gutierrez, Ting Fang Tan, and Daniel Shu Wei Ting. "Large language models in medicine." Nature medicine 29, no. 8 (2023): 1930-1940.
\bibitem{25} Zhang, Shengyu, Linfeng Dong, Xiaoya Li, Sen Zhang, Xiaofei Sun, Shuhe Wang, Jiwei Li et al. "Instruction tuning for large language models: A survey." arXiv preprint arXiv:2308.10792 (2023).
\bibitem{26} Pahalavan Rajkumar Dheivanayahi. “GitHub - Rdpahalavan/PADEC: Thesis Research on Enhancing Network Intrusion Detection System (NIDS) Explainability Using Transformers.” GitHub, n.d. https://github.com/rdpahalavan/PADEC.
\bibitem{27} meta-llama/Llama-2-7b · Hugging Face. “Meta-Llama/Llama-2-7b · Hugging Face,” n.d. \url{https://huggingface.co/meta-llama/Llama-2-7b}.
\bibitem{28} tiiuae/falcon-7b · Hugging Face. “Tiiuae/Falcon-7b · Hugging Face,” June 20, 2023. \url{https://huggingface.co/tiiuae/falcon-7b}.
\bibitem{29} AI, Mistral. “Mixtral of Experts.” Mixtral of experts | Mistral AI | Open-weight models, December 11, 2023. \url{https://mistral.ai/news/mixtral-of-experts/}.
\bibitem{30}  Jiang, Albert Q., Alexandre Sablayrolles, Antoine Roux, Arthur Mensch, Blanche Savary, Chris Bamford, Devendra Singh Chaplot et al. "Mixtral of experts." arXiv preprint arXiv:2401.04088 (2024).

\bibitem{31} Peuster, Manuel, Johannes Kampmeyer, and Holger Karl. "Containernet 2.0: A rapid prototyping platform for hybrid service function chains." In 2018 4th IEEE Conference on Network Softwarization and Workshops (NetSoft), pp. 335-337. IEEE, 2018.
\bibitem{32} Home | TCPDUMP \& LIBPCAP. "Home | TCPDUMP \& LIBPCAP," n.d. \url{https://www.tcpdump.org/}.
\bibitem{33}  Bawany, Narmeen Zakaria, Jawwad A. Shamsi, and Khaled Salah. "DDoS attack detection and mitigation using SDN: methods, practices, and solutions." Arabian Journal for Science and Engineering 42 (2017): 425-441.
\end{thebibliography}

\end{document}